%% file: 0_full_paper.tex
\documentclass[letterpaper]{article}

\usepackage{aaai}
\usepackage{times}
\usepackage{helvet}
\usepackage{courier}
\usepackage{graphicx}

\usepackage{booktabs}
\usepackage{amsmath}
\usepackage{amssymb}
\usepackage{tabularx}
\usepackage{xspace}
\usepackage[noend]{algpseudocode}
\usepackage[T1]{fontenc}
\usepackage{color}
\usepackage{comment}
\usepackage{multirow}
\usepackage[normalem]{ulem}
\usepackage[flushleft]{threeparttable}
\usepackage{fancyvrb}
\usepackage{pdfpages}
\usepackage{url}
\usepackage{caption}
\usepackage{subcaption}
\usepackage{enumitem}
\usepackage[ruled]{algorithm2e} 

\graphicspath{{Figures/}}
\newcommand{\tool}{\textsc{TwiRole}\xspace}

\nocopyright

\begin{document}
%
\title{A Hybrid Model for Role-related User Classification on Twitter}
\author{Liuqing Li \quad Ziqian Song \quad Xuan Zhang \quad Edward A. Fox \\ \\
Department of Computer Science, Virginia Tech \\
Blacksburg, VA 24061, USA \\
\{liuqing, ziqian, xuancs, fox\}@vt.edu
}
\maketitle
\begin{abstract}
\begin{quote}
To aid a variety of research studies, we propose \tool{}, a hybrid model for \emph{role-related user classification on Twitter}, which detects male-related, female-related, and brand-related (i.e., organization or institution) users.
\tool{} leverages features from tweet contents, user profiles, and profile images, and then applies our hybrid model to identify a user's role. 
To evaluate it, we used two existing large datasets about Twitter users, and conducted \emph{both intra- and inter-comparison experiments}. 
\tool{} outperforms existing methods and obtains more balanced results over the several roles. 
We also confirm that user names and profile images are good indicators for this task.
Our research extends prior work that does not consider brand-related users, and is an aid to future evaluation efforts relative to investigations that rely upon self-labeled datasets.
\end{quote}
\end{abstract}

\maketitle
\thispagestyle{empty}

\input{1_introduction}

\input{2_relatedwork}

\input{3_data}

\input{4_modeldesign}

\input{5_experiments}

\input{6_discussion}

\input{7_conclusion}

\section{Acknowledgements}
We thank anonymous reviewers for their thorough comments on this manuscript. 
We thank XXX for funding the XXX project.

\bibliography{aaai_bib}
\bibliographystyle{aaai}
\end{document}

%% file: 1_introduction.tex
\section{Introduction}
Discovering the roles of users in social media such as Twitter, Facebook, and YouTube has attracted people's attention for a decade. 
Accurate classification of users as to role can be helpful in both academic and industrial research. 
Social scientists can undertake more user-centered research, while consumer-oriented companies can provide targeted services.
Accordingly, we seek to identify gender-based roles, as well as a role we refer to as ``brand'', indicating an origin with a company or organization.

Due to privacy concerns, a user's role may not be explicitly revealed.
For example, with Twitter, the gender value can be set by user or predicted by Twitter on the profile page, but that is unreadable to others. 
On the other hand, since Twitter can be helpful in marketing, it is being used by a huge number of brands for product advertising and social influence. 
In 2016, 65.8\% of U.S. companies with 100 or more employees used Twitter for marketing purposes.
Further, on average, each Twitter user follows five brands \cite{Brandwatch}. Thus, it is desirable to have an efficient and accurate model to identify the roles of Twitter users.

Users on Twitter exhibit particular behaviors, and play different roles. 
According to previous research, male users prefer talking about technology and sports \cite{bamman2014gender} while female users tend to show their emotions \cite{rao2010classifying}.
A brand, identified as neither male nor female, is likely to provide job offers or publish advertisements. 
Therefore, we divide the roles of users on Twitter into three categories: \textbf{male-related}, \textbf{female-related}, and \textbf{brand-related}. 
For the role of a trans-gender user, we focus on social gender, since it can best describe her/his behavior. 
Thus, if a female has a social gender identity of male, ``she'' will be considered as a male-related user, and vice versa.

In this paper, we propose a hybrid model to identify different role-related users on Twitter. 
The novel model is consisted of four components, including a basic feature (BF) multi-classifier, an advanced feature (AF) multi-classifier, and a convolutional neural network (CNN) model for profile images. 
Then, considering the output of the three components as an input, we apply another multi-classifier to predict the role of each user. 
In summary, we have made the following contributions:

\begin{itemize}
\item We designed and implemented \tool{}, a hybrid model for multi-role classification. \tool{} achieves better performance by using improved methods (e.g., name parsing), novel features (e.g., first-person words, brightness), and a mix of classifiers. 
\item We conducted an empirical study to compare our model with other popular approaches on the same datasets. \tool{} outperforms them in term of accuracy and obtains more balanced results.
\item We investigated the sensitivity of \tool{} by changing the configurations, and also made two observations. Female-related users are likely to post self-related tweets, and brand-related users prefer to use brighter profile images.  
\end{itemize}

The rest of this paper is structured as follows. 
We discuss the previous research work in Section \ref{Related Work}, and describe our data in Section \ref{Data}. 
In Section \ref{Model} we describe our hybrid model in detail. 
We demonstrate our experimental results in Section \ref{Experiments}, discuss the findings and limitations of \tool{} in Section \ref{Discussion}, and conclude in Section \ref{Conclusion}.

%% file: 2_relatedwork.tex
\section{Related Work}\label{Related Work}

Currently, there is a large body of work on role-related user classification on Twitter, which is usually referred to as gender classification. 
Targeting this bi-classification problem, researchers take user profiles into account, which provides multiple features for classification. 
\cite{Liu2013} carried out the first thorough investigation of the link between gender and first name, and considered the first name as an important feature in gender inference. 
Some researchers used color-based features (e.g., background color, sidebar color, sidebar color) for pre-study and gender prediction \cite{alowibdi2013empirical,alowibdi2013language,burger2011discriminating,ferrarigender,fortmann2013effects}. 
Descriptions of users, i.e., sketches, have been studied, since they may indicate the user gender through gender-based words (e.g., man, woman, boy, girl) \cite{burger2011discriminating,daas2016profiling,pennacchiotti2011machine,vicente2015twitter}. 
Considering different datasets and word dictionaries, \cite{pennacchiotti2011machine} found that 80\% of Twitter users employed such words, but resulting predictions had low accuracy.
On the other hand, \cite{burger2011discriminating} discovered 15\% of the sample users had such an explicit gender cue. 
A user's immediate network (e.g., number of followers/friends) and communication behavior (e.g., retweet frequency) can also be leveraged for classification \cite{ciot2013gender,lasorsa2012transparency,Liu2013,nilizadeh2016twitter,pennacchiotti2011machine,rao2010classifying}. 
Some external sources like a user's website or Facebook page can be helpful in identifying gender too \cite{burger2011discriminating}.
Based on the popular scale invariant feature transformation (SIFT) published in \cite{lowe1999object}, \cite{chen2015comparative} explored features extracted from profile images.

As a key element, tweet contents are worthy of further investigation. Basically, special terms like entities, mentions, and hashtags have been widely used \cite{artwick2014news,bamman2014gender,bergsma2013broadly,ciot2013gender,cunha2012gender,fink2012inferring,li2014weakly,nguyen2013old,pennacchiotti2011machine,geng2017soft}. 
Bag-of-words is another popular feature that helps to differentiate the k-top words in different user groups \cite{bamman2014gender,Liu2013,pennacchiotti2011machine}. 
There is a conclusion that females tend to use more emotional words, while males tend to use more numbers and technology words \cite{rao2010classifying,bamman2014gender}. 
Furthermore, some researchers mainly focused on tweets and applied n-gram character features to detect the gender of users \cite{al2012homophily,burger2011discriminating,ciot2013gender,deitrick2012gender,liu2012using,miller2012gender,rao2010classifying}. 
Some rich linguistic features are also generated from unsupervised learning methods like Latent Dirichlet Allocation (LDA) \cite{pennacchiotti2011machine,ramage2010characterizing,ferrarigender,geng2017soft}. 
Image sets posted by Twitter users provide another type of source that has been further developed \cite{sakaki2014twitter,ma2014gender}. 

Additionally, deep neural networks have been leveraged in gender classification on social media in recent years. 
Various CNN models are designed to classify the gender in social media \cite{levi2015age,wang2016voting,wang2016deciphering}. 
However, the dataset of \cite{levi2015age} is produced from Flickr instead of Twitter, while \cite{wang2016voting,wang2016deciphering} only pay attention to profile images with faces. \cite{geng2017soft} proposed an ensemble approach by combining models for both tweet contents and profile images to improve the quality of bi-classification.

However, a crowd-sourcing experiment designed by Nguyen et al. \cite{Nguyen2014} discovered that user prediction based only on tweet contents seems a bit difficult because of the difference between gender and gender identity. 
Besides this, there are methodological concerns regarding the evaluation of studies, since most researchers evaluate their approaches on self-labeled datasets instead of open datasets or benchmarks \cite{Liu2013}. 
Further, it is not enough to categorize users only into male or female, since there is still a certain proportion of brand-related users, as we mentioned above. 
To the best of our knowledge, the work of Ferrari et al. \cite{ferrarigender} is the only which studied the three category user classification problem, which is most relevant to our exploration. 
They trained a gradient boosting classifier to generate predictions resulting from traditional features (e.g., bag-of-words, tf-idf, topics) to identify different role-related users. 
While applying several simple features, \tool{} also makes use of some novel features such as first-person words in tweets, brightness in profile images, term frequency in descriptions, etc.

%% file: 3_data.tex
\section{Data}\label{Data}
To reduce bias resulting from the selection of data, we did not create any labeled dataset by ourselves. 
Instead, we reused two existing Twitter user classification datasets, each of which contains a huge number of users with labels. 
In this section, we first describe each dataset (Section \ref{Description}) and then discuss the preprocessing (Section \ref{Preprocessing}).

\begin{table*}[ht]
\small
\centering
\caption{Data Preprocessing on Kaggle Dataset}
\label{tab:preprocessing}
\begin{tabular}{|c|l|c|}
\hline
\textbf{Step} & \multicolumn{1}{c|}{\textbf{Action}}       & \textbf{Number of Users Left}                      \\ \hline
0    & \multicolumn{1}{c|}{----}                  & 20,050                                        \\ \hline
1    & Remove duplicated users                    & 18,795                                        \\ \hline
2    & Remove users with blank and ``unknown'' labels & 17,660                                        \\ \hline
3    & Remove users with less confidence value    & 12,991                                        \\ \hline
4    & Remove robot-like users                    & 12,889                                        \\ \hline
5    & Remove users with tweet file size < 4KB    & 8,714                                         \\ \hline
6    & Remove users with broken profile images    & 8,625 (male: 3,195, female: 3,176, brand: 2,254) \\ \hline
7    & Subsampling                                & 6,000 (male: 2,000, female: 2,000, brand: 2,000) \\ \hline
\end{tabular}
\end{table*}

\subsection{Description}\label{Description}

\begin{enumerate}
\item \textit{Kaggle Dataset}
\end{enumerate}
The dataset on Kaggle \cite{Kaggle} is a project of CrowdFlower \cite{CrowdFlower}, including the information of about 20,000 users.
Project contributors manually labeled each user by checking the corresponding information, which contains part of the profile metadata, such as display name, screen name, description, link color, etc. 
There are three labels in the dataset: male, female, and brand. 
The contributors also provided a confidence score along with the role tag, which is a good indicator of labeling quality.
\begin{enumerate}[resume]
\item \textit{Gender-labeled Twitter Dataset}
\end{enumerate}
Liu and Ruths released a public gender-labeled dataset\footnote{Download link: http://www.networkdynamics.org/static/datas-ets/LiuRuthsMicrotext.zip} to support the evaluation of different user detection approaches. 
Three Amazon Mechanical Turk workers manually labeled a user as male or female if all of them agreed on the same gender assignment. The dehydrated dataset only has two fields: user ID and gender; the gender field has two values: ``M'' and ``F''. 
In total, there are 4,449 male-related users and 8,232 female-related users.

\subsection{Preprocessing}\label{Preprocessing}
For \textit{Dataset 1}, we first removed duplicate users that have the same screen name.
Since we focused on role-related users, those with blank labels or ``unknown'' labels were also filtered out. 
We kept users with labels having confidence value 1, assuming their records are of high labeling quality. 
Regular expressions were utilized to detect and remove users likely to be robots. 
For the remaining users, we ran a 24 virtual machine cluster to retrieve their tweets and eliminated any user whose tweet file size is less than 4KB. 
Next, we updated the user profiles and crawled the ``bigger'' profile image of each user through the Twitter API.
Thus, we created a high quality Twitter user dataset that contains profile information, tweet contents, and profile images. 
The resulting 8,625 high quality set of users had unbalanced sizes among the different roles. 
Consequently, we randomly selected 6,000 users (2,000 users in each category) to build a subsampled dataset. 
Table \ref{tab:preprocessing} shows all of the steps, giving the number of users at each step.
%
%

For \textit{Dataset 2}, we followed step 5-7 to retrieve the profiles, tweets and images of users, and completed the preprocessing task. It is noticeable that the gender-labeled Twitter dataset only has two classes, then we took 3,000 users as a subset of each class. 
%
%

%% file: 4_modeldesign.tex
\section{Model Design}\label{Model}

\begin{figure*}[ht]
    \centering
    \includegraphics[width=.8\textwidth]{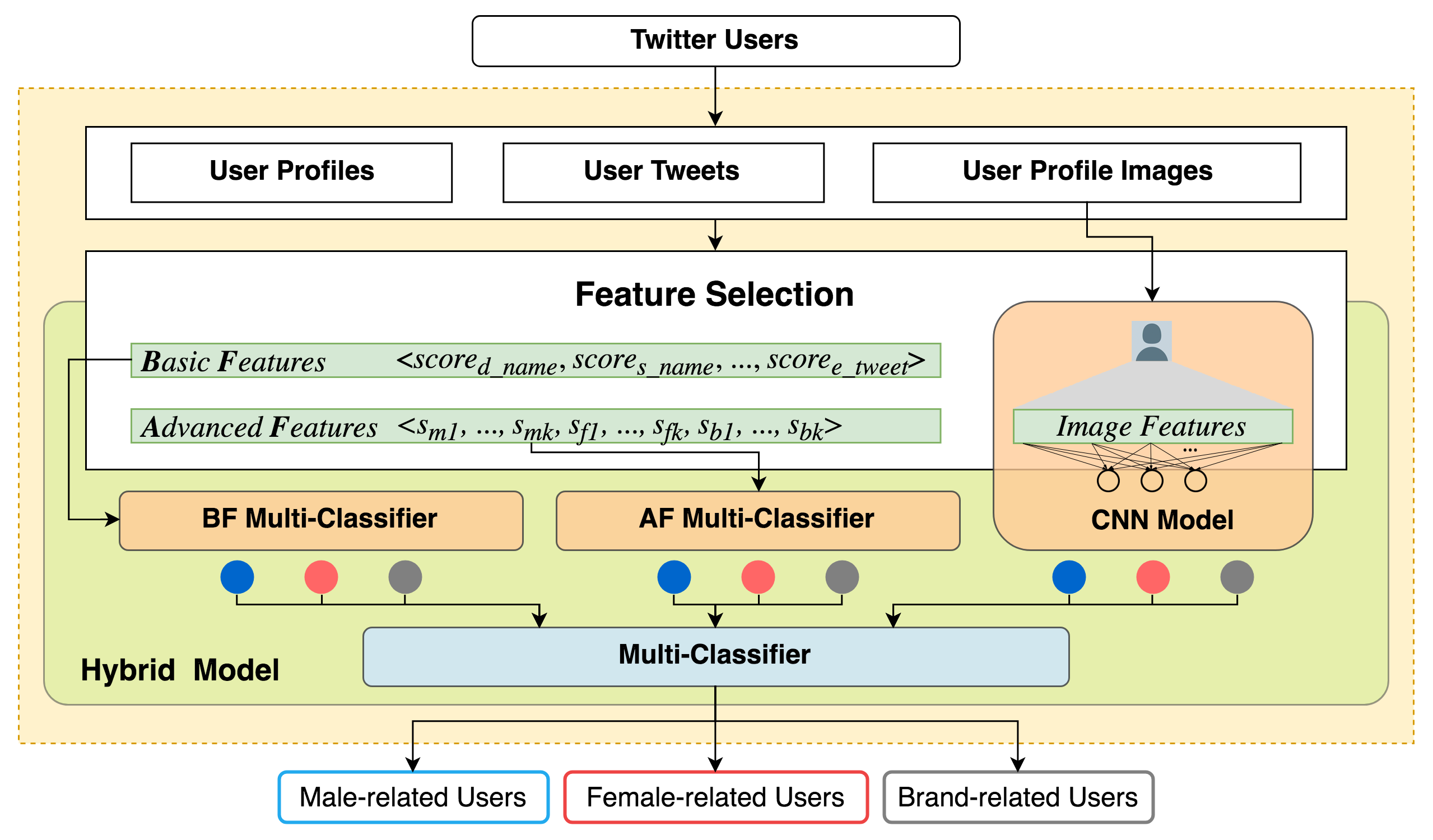}
    \caption{\tool{} is a hybrid model consisting of four components. 
The BF multi-classifier takes the basic features from the user profiles and tweets as an input. 
The AF multi-classifier focuses on the k-top words in user tweets. 
CNN works on the user profile images. 
The final multi-classifier takes the output of the above three modules as an input during training and testing.}
    \label{fig:architecture}
%
%
\end{figure*}

We designed and implemented \tool{} to classify the role-related users. 
This hybrid model has a BF multi-classifier, an AF multi-classifier, a CNN model, and a final multi-classifier, as shown in Fig. \ref{fig:architecture}. 
There are training and testing phases in our approach. 
Both phases extract the features from user profiles, tweets, and images, which characterize the role of users. 
In this section, we will first describe the feature selection and calculation (Section \ref{Feature Selection and Calculation}), and then explain the training and testing phases (Section \ref{Training and Testing}).

\subsection{Feature Selection and Calculation}\label{Feature Selection and Calculation}
We mainly focus on the following five types of features: \textit{name}, \textit{description}, \textit{relationship}, \textit{profile image}, and \textit{tweet}. 
Each type has one or multiple features, as shown in Table \ref{tab:feature_type}, and different modules are focused on different features.

\begin{table}[h]
\small
\centering
\caption{Feature Types and Details in \tool{}}
\label{tab:feature_type}
\begin{tabular}{|c|c|l|}
\hline
\textbf{No.}         & \textbf{Feature Type}        & \textbf{Feature Detail} \\ \hline
\multirow{2}{*}{BF1} & \multirow{2}{*}{name}        & display name            \\ \cline{3-3} 
                     &                              & screen name             \\ \hline
\multirow{2}{*}{BF2} & \multirow{2}{*}{description} & first-person score      \\ \cline{3-3} 
                     &                              & term frequency          \\ \hline
BF3                  & relationship                 & TFF score               \\ \hline
BF4                  & profile image                & brightness              \\ \hline
\multirow{3}{*}{BF5} & \multirow{3}{*}{tweet}       & first-person score      \\ \cline{3-3} 
                     &                              & interjection score      \\ \cline{3-3} 
                     &                              & emotion score           \\ \hline
AF1                  & tweet                        & k-top words             \\ \hline
CNN1                 & profile image                & hidden in image         \\ \hline
\end{tabular}
\end{table}

\begin{enumerate}
\item \textit{BF1 -- name}
\end{enumerate}
To calculate the name score, we downloaded popular baby names  from the US Social Security Administration \cite{babynames} into a database, chose a subset covering the past 10 years, and then summarized the occurrences of names over years. 
Each name can be represented as a vector $<name, gender, frequency>$.
We had a total of 71,299 records. 
It must be noted that for a given name, the gender could be either male or female. 
For instance, there are 406 female babies named ``dallis'' while 167 male babies have the same name. 
To expand the name dataset, we combined it with an Arabic name dictionary \cite{arabicnames} that includes 979 female names and 898 male names. 
Because the Arabic dataset has no occurrence numbers, we simply set the field with the same value for each name. 
Duplicated names were removed during combination.
Finally there were 72,134 rows in the name dictionary.

Given a display name $d\_name$, \tool{} first tokenizes it into terms.
It only takes the term that first appears in the name dictionary to calculate the display name score $score_{d\_name}$.
If there is no term found, $score_{d\_name}$ is equal to 0, as shown in Equation \ref{eq:name_d}:

\begin{equation}
\small
\label{eq:name_d}
score_{d\_name}= 
\begin{cases}
	\frac { tf_f-tf_m}{ \max {(tf_f,tf_m)}} \in [-1,1] ,& \text{term $t$ is found}\\
    0,              & \text{otherwise}
\end{cases}
\end{equation}
where $tf_f$ and $tf_m$ represent the female and male frequency of term $t$. For instance, given a display name ``John Clemson'', since ``John'' is the first term found in the name dictionary and there are 445 females and 256,166 males named ``John'', the display name score is calculated as $score_{d\_name} = (445 - 256166) / 256166 = -0.998$.

For the screen name, \tool{} parses the entire string into terms through an integrated method combined with four different methods: name-based, word-based, name-word-based, and wordninja \cite{wordninja}, a popular word splitter. 
Based on the parsing results, \tool{} takes the result with the least number of terms as the best candidate. Table \ref{tab:name_s} shows the different parsing results of two samples, where the candidates are shown in bold. 
Then, we reapply Equation \ref{eq:name_d} to calculate the screen name score $score_{s\_name}$.

\begin{table}[h]
\small
\centering
\caption{Results for Different Parsing Methods}
\label{tab:name_s}
\begin{tabular}{|c|c|c|}
\hline
\textbf{screen name} & clemsonjohn            & 123tommy               \\ \hline
\textbf{Method}      & \multicolumn{2}{c|}{\textbf{Results}}           \\ \hline
name-based           & clem, son, john        & tom, my                \\ \hline
word-based          & cl, ems, on, john      & \textbf{tommy}         \\ \hline
name-word-based      & clem, son, john        & \textbf{tommy}         \\ \hline
wordninja            & \textbf{clemson, john} & 1, 2, 3, t, o, m, m, y \\ \hline
\end{tabular}
\end{table}

\begin{enumerate}[resume]
\item \textit{BF2 -- description}
\end{enumerate}
Users on Twitter are likely to show their role information through the description, since it appears on the personal main page, and might give a brief introduction to the user. 
We proposed two word lists: first-person word list and brand word list, to calculate the first-person score in the description. 
The first list is represented as $list_{first}$, containing first-person words like $i, am, my, me, mine, i'm$, while the latter one is represented as $list_{brand}$, which has one word: \textit{official}. 
While scanning the terms in a user's description, we set the first-person score by following the rules below.
%
%

\begin{equation}
\small
\label{eq:name_s}
score_{fp\_desc}= 
\begin{cases}
	1,		& \text{term $t \in list_{first}$ and $t \notin list_{brand}$}\\
    -1,		& \text{term $t \in list_{brand}$ and $t \notin list_{first}$} \\
    0,		& \text{otherwise}		
\end{cases}
\end{equation}
Then, we removed hashtags, mentions, and URLs from the user's description, and stored the occurrence of the remaining terms for each user as $score_{tf\_desc}$.

\begin{enumerate}[resume]
\item \textit{BF3 -- relationship}
\end{enumerate}
Twitter Follower-Friend (TFF) ratio is a widely used figure that represents the quantitative relation of a user \cite{krishnamurthy2008few}. 
Here, we make use of the number of followers $Num_{followers}$ and friends $Num_{friends}$ to calculate the TFF ratio, as shown in Equation \ref{eq:tff}:

\begin{equation}
\small
\label{eq:tff}
score_{tff}= \log{\frac{{Num_{followers}}^{2}+1}{Num_{friends}+1}}
\end{equation}
Different from the basic TFF score, we add 1 to both denominator and numerator to avoid dividing by zero or $\log{0}$ error. 
We also add an exponent on the numerator to strengthen the number of followers and differentiate the cases where the number of followers is proportional to the number of friends.

\begin{enumerate}[resume]
\item \textit{BF4 -- profile image}
\end{enumerate}
For the profile images, \tool{} focuses on the HSV format instead of RGB, because the latter format seems more noisy. We selected brightness, also called ``value'' in HSV, as a basic feature. Given a profile image, we convert the original RGB format into HSV, accumulate the brightness score of each pixel, and compute the average brightness $score_{b\_image}$ of the entire image.

\begin{enumerate}[resume]
\item \textit{BF5 -- tweet}
\end{enumerate}
To process the tweet contents of each user, \tool{} generates three scores: first-person score, interjection score, and emotion score, which are represented as $score_{fp\_tweet}$, $score_{i\_tweet}$, and $score_{e\_tweet}$, respectively. 
Besides the first-person word list above, we created two other word lists: interjection list and emotion word list\footnote{The two lists can be found at the following link: \textbf{(not disclosed for anonymous review)}} for the term matching. 
The three scores can be calculated through a linear scan of each tweet collection. 
Equation \ref{eq:tweet_basic} shows how to calculate the first-person score; the other two scores can be computed in the same way.

\begin{equation}
\small
\label{eq:tweet_basic}
score_{fp\_tweet}= \frac{\text{\# of tweets that have terms in $list_{first}$}}{\text{\# of tweets in user tweet collection}}
\end{equation}

\begin{enumerate}[resume]
\item \textit{AF1 -- tweet}
\end{enumerate}
\tool{} follows the popular k-top words method \cite{Liu2013} to further process each tweet collection. 
First, TweetNLP \cite{owoputi2013improved}, a fast and robust tokenizer and part-of-speech tagger, has been leveraged to extract the important tags -- nouns (N), verbs (V), adjectives (A), adverbs (R), emoticons (E) and hashtags (\#) -- from the raw texts. 
Then, in comparison with the k-top words method, our tool applies a similar method to create a word list for each of the three roles and merge them into one vector, represented as $vector_{k\_top}=<w_{m1}, w_{m2}, ..., w_{mk}, w_{f1}, w_{f2}, ..., w_{fk}, w_{b1}, w_{b2}, ..., w_{bk}>$. 
For each word in the vector, \tool{} counts the number of tweets containing the given word, divides it by the size of the user tweet collection, and finally gets the k-top words score vector $score_{k\_top}$, shown in Equation \ref{eq:tweet_adv}. 
By default, we set $k=20$ in our approach. 
\begin{equation}
\small
\label{eq:tweet_adv}
\begin{split}
score_{k\_top} = <&s_{m1}, s_{m2}, ..., s_{mk}, \\
& s_{f1}, s_{f2}, ..., s_{fk}, s_{b1}, s_{b2}, ..., s_{bk} >
\end{split}
\end{equation}

\begin{enumerate}[resume]
\item \textit{CNN1 -- profile image}
\end{enumerate}
In addition to the basic and advanced features, our tool also applies a pre-trained ResNet-18 model \cite{he2016deep} to extract the hidden features from the profile images. 
Because the original ResNet is designed for the ImageNet dataset \cite{russakovsky2015imagenet} with 1,000 categories, we have changed the number of nodes in the output layer from 1,000 to 3 and employed a softmax function on the three nodes. 
There are 512 nodes in the fully-connected layer which produce the deep features in the network. 

\subsection{Training and Testing}\label{Training and Testing}
We utilize 10-fold cross validation to evaluate \tool{}. 
In the training phase, we calculate all the feature scores from BF1 to BF5 for the users as input, and train the BF multi-classifier with the role-related labels. 
For each user, the output is a probability vector of three different roles. 
Then, we train the AF multi-classifier with the k-top words score vectors in the same way. 
Next, the profile images of all the training users and their labels are put into the ResNet-18 model to train the deep neural network, and we can also get the probability vector for each user. 
At last, we concatenate the three probability vectors and train the final multi-classifier.
Different types of classifiers (e.g., decision tree, Naive Bayes) can be applied on the AF, BF, and final multi-classifiers. To reduce the number of combinations, we set all the three multi-classifiers with the same type. 
In the testing phase, since all the modules have been trained and fixed, for each user, we just follow the above steps to produce the prediction from the final multi-classifier and compare the result with the ground truth.

%% file: 5_experiments.tex
\section{Experiments}\label{Experiments}
In this section, we evaluate \tool{} on two datasets. For the Kaggle dataset, we carry out an intra-comparison to verify our tool with different classifiers, features, and parameters, and also draw an inter-comparison between \tool{} and the method developed by Ferrari et al. \cite{ferrarigender}. 
Then, for the gender-labeled Twitter dataset, we slightly modify \tool{} into a bi-classification model \tool{}$^{bi}$, and compare it with Liu \& Ruths' approach on their dataset. 
The metrics used in our experiments will be introduced in Section \ref{Metrics}, and the detailed results are described in Section \ref{Results}.

\subsection{Metrics}\label{Metrics}
We use a confusion matrix to calculate the recall (R), precision (P), and F1 score of each role. 
For a certain role $r$, the three values are computed as:
\begin{equation}
\small
\label{eq:rec_pre}
\begin{aligned}
Recall_r &= \frac{\text{\# of users correctly identified as } r}{\text{\# of users labeled as }r}, \\
Precision_r &= \frac{\text{\# of users correctly identified as } r}{\text{\# of users predicted as } r}, \\
F1_r &= \frac{2 * Recall_r * Precision_r}{Recall_r + Precision_r}
\end{aligned}
\end{equation}
The performance of \tool{} is reflected in the overall accuracy; see Equation \ref{eq:acc}.

\begin{equation}
\small
\label{eq:acc}
Accuracy = \frac{\sum_{r}{(\text{\# of users correctly identified as } r)}}{\sum_{r}{(\text{\# of users labeled as }r)}}
\end{equation}

\subsection{Results}\label{Results}
\begin{enumerate}
\item \textit{Kaggle Dataset}
\end{enumerate}
First, we evaluate \tool{} with different classifiers and also measure the performance of each single model as well as our hybrid model. 
Regarding a multi-classifier that considers the basic and advanced features (BF, AF), we experiment to compare classical individual classifiers like decision tree and support vector machine (SVM), and ensemble classifiers such as AdaBoost, GradientBoosting, and random forest. 
For the CNN model, we use ResNet-18 as default. 

\begin{table*}[ht]
\small
\centering
\caption{Accuracy of \tool{}'s modules with different classifiers}
\label{tab:classifier}
\begin{tabular}{|c|c|c|c|c|}
\hline
\multirow{2}{*}{\textbf{Classifier Type}} & \multicolumn{4}{c|}{\textbf{Accuracy}} \\ \cline{2-5}
& \textbf{BF Multi-classifier} & \textbf{AF Multi-classifier} & \textbf{Profile Image CNN} & \textbf{Overall} \\ \hline
Decision Tree 		& 0.721 & 0.618 & \multirow{5}{*}{0.790} & 0.721 \\ \cline{1-3} \cline{5-5} 
SVM          		& 0.739 & 0.352 & & 0.800 \\ \cline{1-3} \cline{5-5} 
AdaBoost     		& 0.790 & 0.704 & & 0.850 \\ \cline{1-3} \cline{5-5} 
GradientBoosting	& \textbf{0.816} & \textbf{0.738} & & 0.842 \\ \cline{1-3} \cline{5-5} 
Random Forest		& 0.796 & 0.708 & & \textbf{0.899}\\ \hline
\end{tabular}
\end{table*}

\begin{table*}[ht]
\small
\centering
\caption{\tool's performance with different feature sets}
\label{tab:feature}
\begin{tabular}{l | c c c | c c c | c c c | c}
\toprule
\multirow{2}{*}{\textbf{Feature Set Description}} & \multicolumn{3}{c|}{\textbf{Male}} & \multicolumn{3}{c|}{\textbf{Female}} & \multicolumn{3}{c|}{\textbf{Brand}} & \multirow{2}{*}{\textbf{Acc}} \\ \cline{2-10}
& \textbf{R} & \textbf{P} & \textbf{F1} & \textbf{R} & \textbf{P} & \textbf{F1} & \textbf{R} & \textbf{P} & \textbf{F1} \\ \hline
0. All Features & 0.885 & 0.922 & \textbf{0.903} & 0.920 & 0.897 & \textbf{0.908} & 0.891 & 0.879 & \textbf{0.885} & \textbf{0.899} \\ \hline
1. Without BF1 (name) & 0.845 & 0.906 & 0.874 & 0.878 & 0.874 & 0.876 & 0.888 & 0.837 & 0.861 & 0.870 \\ \hline
2. Without BF2 (description) & 0.881 & 0.928 & 0.903 & 0.916 & 0.895 & 0.905 & 0.895 & 0.873 & 0.883 & 0.897 \\ \hline
3. Without BF3 (relationship) & 0.878 & 0.926 & 0.901 & 0.919 & 0.893 & 0.906 & 0.892 & 0.873 & 0.882 & 0.896 \\ \hline
4. Without BF4 (profile image) & 0.875 & 0.922 & 0.897 & 0.914 & 0.892 & 0.902 & 0.888 & 0.865 & 0.876 & 0.892 \\ \hline
5. Without BF5 (tweet) & 0.881 & 0.924 & 0.901 & 0.904 & 0.892 & 0.898 & 0.890 & 0.863 & 0.875 & 0.892 \\ \hline
6. Without AF1 (tweet) & 0.874 & 0.913 & 0.893 & 0.904 & 0.883 & 0.893 & 0.876 & 0.862 & 0.868 & 0.885 \\ \hline
7. Without CNN1 (profile image) & 0.797 & 0.834 & 0.814 & 0.860 & 0.828 & 0.843 & 0.856 & 0.855 & 0.854 & 0.837 \\ \bottomrule
\end{tabular}
\end{table*}

Table \ref{tab:classifier} shows the accuracy of \tool{}'s modules with different classifiers.
The CNN model alone does well, but a combination is better.
Among the five classifiers, GradientBoosting does best for both set of features ($Acc_{BF} = 0.816$, $Acc_{AF} = 0.738$), but random forest has the highest accuracy ($Acc = 0.899$) regarding the entire model. 
Moreover, the ensemble classifiers perform better than the classical individual classifiers in each single model and in the hybrid model. 
Particularly, the accuracy of SVM with the advanced features is only 0.352, which is just slightly better than random results. 
By comparing the performance of each single model and the hybrid model, we notice that the hybrid model is always better than each single model with different classifiers, except decision tree (with a tie). 
Accordingly, in further evaluation studies, the default is to use random forest, and a hybrid overall model is preferred.

Then, we carry out the evaluation on the feature sets of \tool{} that can help us find out which feature has great impact among the whole feature set. 
Specifically, we take the hybrid model with all features as our baseline method, and remove the features belonging to each feature type step by step to generate multiple feature subsets. Based on the remaining features, we retrain and reevaluate the entire model.

Table \ref{tab:feature} shows \tool{}'s performance with different feature sets. 
Using all features gives the best accuracy overall, as well as the best F1 score for each of the roles.
The CNN features seem most important; omitting them leads to a 6.2\% drop in accuracy. 
Similarly, as a basic feature, name (including display and screen name) also plays an important role among the features. 
On the other hand, some features, like description and relationship, may have a small impact; the overall accuracy only declined 0.2\% and 0.3\%, respectively. 
We also consider how \tool{} does with regard to each of the user roles.
In most cases, we find that $P_{male} > P_{female} > P_{brand}$, $R_{female} > R_{brand} > R_{male}$ and $F1_{female} > F1_{male} > F1_{brand}$. 
The exceptions mainly occur in sets 1 and 7, where the dropped features have great impact on prediction results.

Focusing on the user tweet collections, we further investigate the parameters in \tool{}. 
We first choose the most recent 10, 30, 50, and all tweets posted by each user to calculate the three scores in BF5, then set the value k as 1, 5, 10, and 20 in k-top words. 
Table \ref{tab:params} shows the performance of \tool{} with different parameters. 
The accuracy achieves the best result in BF5 when we leverage the entire tweet collection, since it can integrally describe the user's behavior. 
For the k-top words, the best value of k is 10 or 20. 
It seems that a small k is not helpful enough to differentiate the roles of users. 
When we set k to 30, 50, or 100, computation time increases, but with no significant performance increase.

\begin{table}[ht]
\small
\caption{\tool{}'s performance with different parameters in BF5 and AF1}
\label{tab:params}
\parbox{\linewidth}{
\centering
\begin{tabular}{c|c}
\toprule
\textbf{Parameters in BF5 (tweet)}&\textbf{Acc}\\
\midrule
Recent 10 tweets & 0.894 \\ \hline
Recent 30 tweets & 0.893 \\ \hline
Recent 50 tweets & 0.893 \\ \hline
All user tweets & \textbf{0.899} \\ \bottomrule
\end{tabular}
}
\\ \\ \\
\parbox{\linewidth}{
\centering
\begin{tabular}{c|c}
\toprule
\textbf{Parameters in AF1 (tweet)}&\textbf{Acc}\\
\midrule
1 top words & 0.886 \\ \hline
5 top words & 0.890 \\ \hline
10 top words & \textbf{0.899} \\ \hline
20 top words & \textbf{0.899} \\ \bottomrule
\end{tabular}
}
\end{table}

\begin{table*}[ht]
\small
\centering
\caption{Classification results of \tool{} and Ferrari et al.'s work}
\label{tab:inter}
\begin{tabular}{c | c c c | c c c | c c c | c}
\toprule
\multirow{2}{*}{\textbf{Tool}} & \multicolumn{3}{c|}{\textbf{Male}} & \multicolumn{3}{c|}{\textbf{Female}} & \multicolumn{3}{c|}{\textbf{Brand}} & \multirow{2}{*}{\textbf{Acc}} \\ \cline{2-10}
& \textbf{R} & \textbf{P} & \textbf{F1} & \textbf{R} & \textbf{P} & \textbf{F1} & \textbf{R} & \textbf{P} & \textbf{F1} \\ \hline
\tool{}					& 0.885 & 0.922 & 0.903 & \textbf{0.920} & \textbf{0.897} & \textbf{0.908} & \textbf{0.891} & \textbf{0.879} & \textbf{0.885} & \textbf{0.899} \\ \hline
Ferrari et al., 2017	& \textbf{0.948} & \textbf{0.946} & \textbf{0.947} & 0.806 & 0.857 & 0.831 & 0.837 & 0.786 & 0.811 & 0.865 \\ \bottomrule
\end{tabular}
\end{table*}

\begin{table*}[ht]
\small
\centering
\caption{Performances of \tool{}$^{bi}$ and Liu \& Ruths' method}
\label{tab:inter_bi}
\begin{tabular}{c | c c c | c c c | c}
\toprule
\multirow{2}{*}{\textbf{Tool}} & \multicolumn{3}{c|}{\textbf{Male}} & \multicolumn{3}{c|}{\textbf{Female}} & \multirow{2}{*}{\textbf{Acc}} \\ \cline{2-7}
& \textbf{R} & \textbf{P} & \textbf{F1} & \textbf{R} & \textbf{P} & \textbf{F1} \\ \hline
\tool{}$^{bi}$ & 0.896 & \textbf{0.901} & 0.898 & 0.901 & \textbf{0.897} & 0.898 & \textbf{0.899} \\ \hline
Liu and Ruths, 2013 & --  & 0.875 & -- & -- & 0.866 & -- & 0.871 \\ \bottomrule
\end{tabular}
\end{table*}

We compare \tool{} with Ferrari et al.'s work on the same dataset. 
The classification results are shown in Table \ref{tab:inter}. 
Their model has an advantage in identifying the male-related users, where the F1 score is 0.947 and ours is 0.903. 
But \tool{} performs better in detecting both female-related ($F1_{female} = 0.908$) and brand-related users ($F1_{brand} = 0.885$), and the overall accuracy ($Acc = 0.899$) is higher than with Ferrari et al.'s approach ($Acc = 0.865$). 
In addition, the prediction results of our model are more balanced across different roles, because the difference in F1 score is only 0.023 in \tool{} while it is 0.136 for Ferrari et al. 

\begin{enumerate}[resume]
\item \textit{Gender-labeled Twitter Dataset}
\end{enumerate}
Besides testing multi-classification, we also test our hybrid model on the gender-labeled Twitter dataset. 
Because the dataset has only two classes -- male and female -- we slightly adjust \tool{} to enable it for bi-classification; we name the variant model as \tool{}$^{bi}$. 
It makes use of the same features as \tool{}, but merges the basic and advanced features together for bi-classification. Moreover, the output of each module has been changed into two classes to fit the data format.

We still apply 10-fold cross validation to train and evaluate \tool{}$^{bi}$. 
Table \ref{tab:inter_bi} shows the performance of \tool{}$^{bi}$ and Liu \& Ruth's method. 
Since there are no recall values in their paper, we are not able to compare the recall and F1 score. 
Based on precision and accuracy, we see that \tool{}$^{bi}$ has better performance than their method in each role ($P_{male} = 0.901, P_{female} = 0.897$) and in the overall evaluation ($Acc = 0.899$).

%% file: 6_discussion.tex
\section{Discussion}\label{Discussion}
In this section, we first present some interesting findings in Section \ref{Relevant Feature}, and then describe the limitations of our current approach in Section \ref{Limitations}.

\subsection{Relevant Feature}\label{Relevant Feature}
\begin{figure}[h]
\centering
\begin{subfigure}{.45\textwidth}
  \centering
  \includegraphics[width=\linewidth]{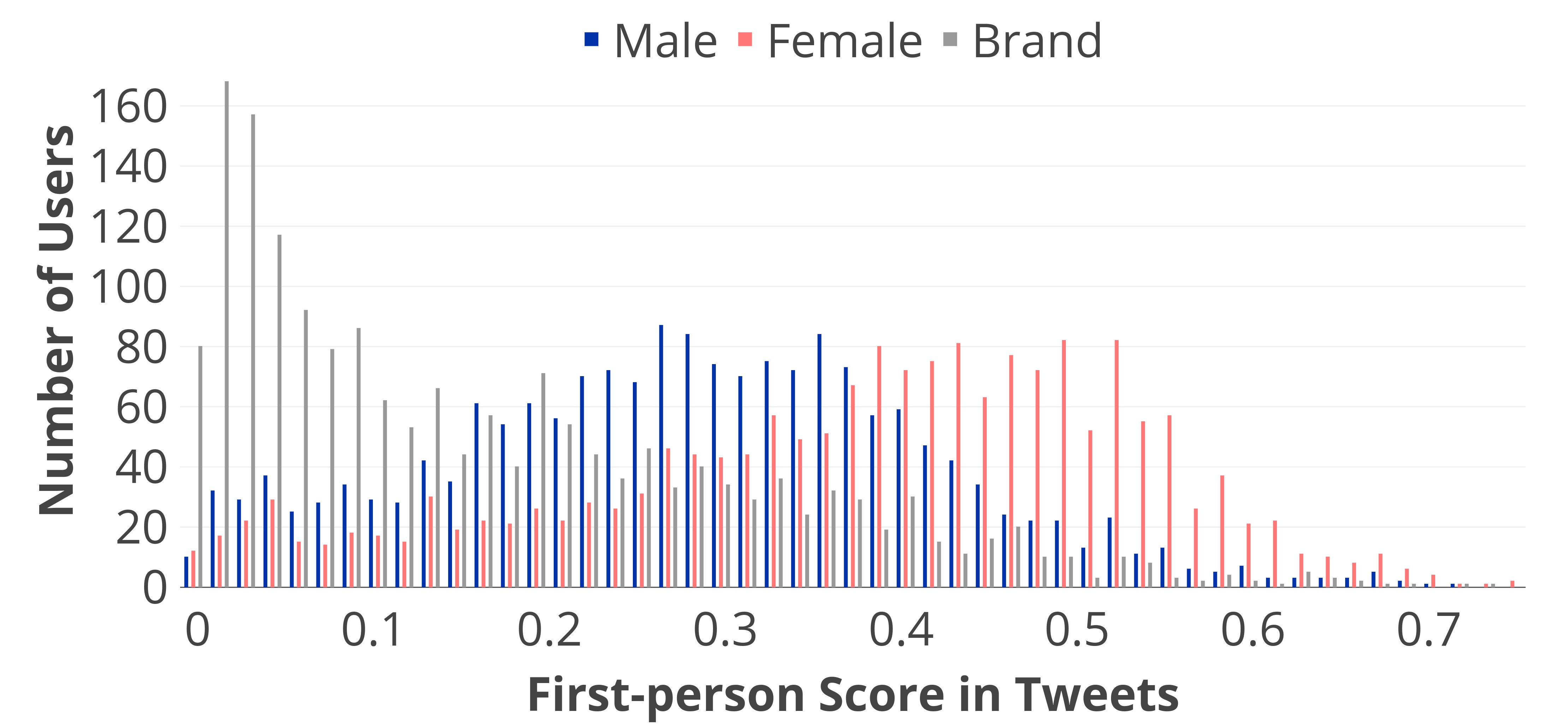}
  \caption{First-person Score Distribution}
  \label{fig:person_score}
\end{subfigure}
\par\bigskip
\begin{subfigure}{.45\textwidth}
  \centering
  \includegraphics[width=\linewidth]{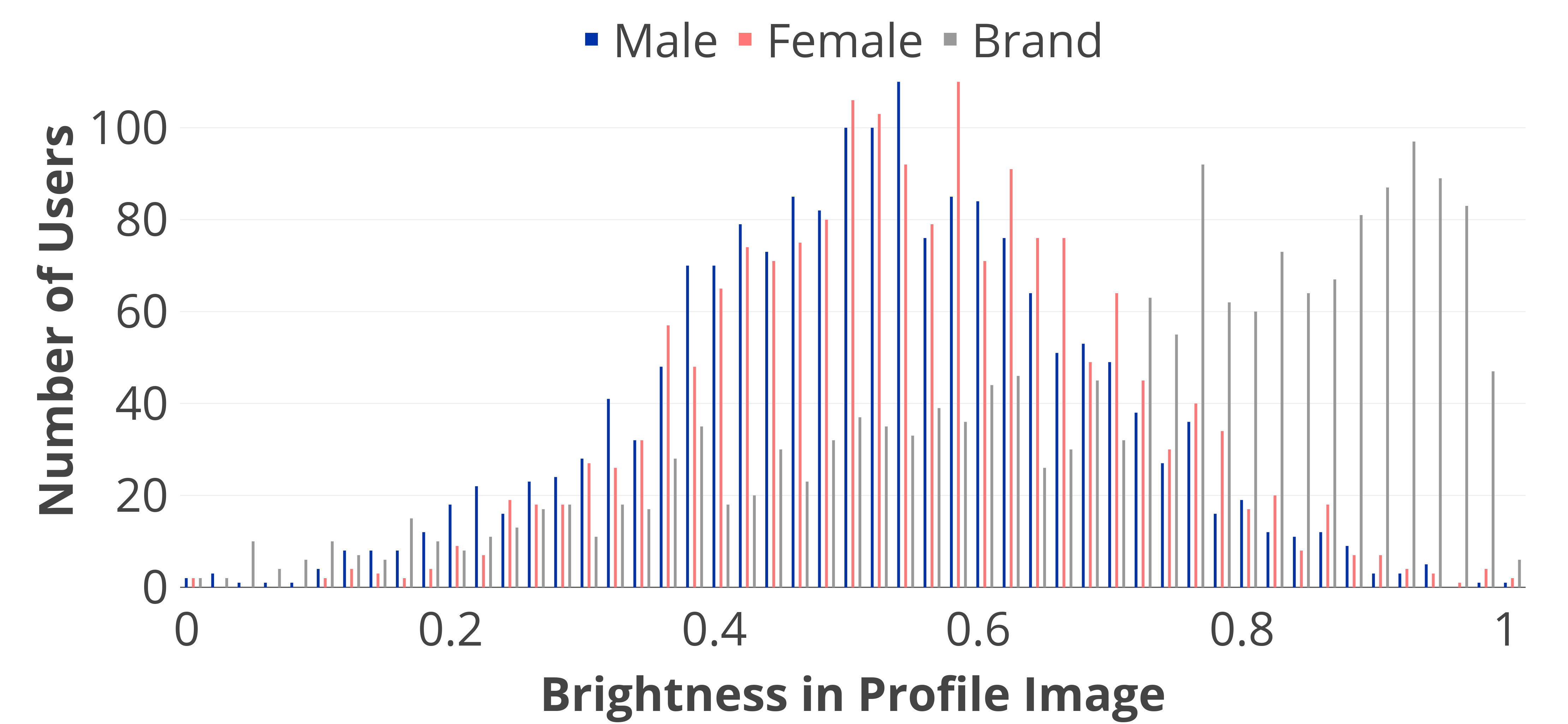}
  \caption{Brightness Distribution}
  \label{fig:brightness}
\end{subfigure}
\caption{Relevant Features Discovered in \tool{}}
\label{fig:features}
\end{figure}

\begin{enumerate}
\item \textit{First-person Score in Tweets}
\end{enumerate}
We take one fold as a sample during the training phase and draw the first-person score distribution, as shown in Fig. \ref{fig:person_score}.
Using a pairwise T-test and requiring p-value < 0.05, we find that each role is statistically significantly different from any other role.
Thus, according to the sample data, brand-related users seldom use first-person words in their tweets, while female-related users are likely to mention themselves through tweets.

\begin{enumerate}[resume]
\item \textit{Brightness in Profile Image}
\end{enumerate}
Similarly, we investigated brightness; see the distribution in Fig. \ref{fig:brightness}. 
The difference between any pair of roles is also significant, even between male-related and female-related users.
The average brightness for female-related users is greater than for male-related users, while brand-related users have even brighter profile images; these may be more engaging.

\subsection{Limitations}\label{Limitations}
One limitation in our study is that \tool{} has been trained and evaluated on only two datasets. 
Though users in the two datasets were randomly selected and manually labeled, and can be considered as good representatives on Twitter, the generality of our approach still needs to be verified through more datasets or practice. 
Second, our name dictionaries are limited to English and Arabic names, which might lead to a language based bias.
Thus, \tool{} might have difficulty identifying some users' names from other sources, including both display name and screen name. 
For instance, if a user's screen name is a male name in Spanish that is not in our dictionary, $score_{s\_name}$ will be calculated as 0 instead of -1. 
Consequently, expanding the dictionary  with Spanish, Chinese, and other types of names should be investigated.
Third, for our CNN model, we simply modified the last layer in a pre-trained ResNet-18 in our user classification task, which led to an accuracy of around 79\% in our main experiment. 
The pre-trained model was originally developed for the ImageNet dataset \cite{russakovsky2015imagenet}, and most of the weights have already been fixed.
Future would could involve optimization to extract more role-related features for our specific task.

%% file: 7_conclusion.tex
\section{Conclusion}\label{Conclusion}
We presented \tool{}, a hybrid model for role-related user classification on Twitter. 
Different from prior work that distinguishes between male and female, \tool{} is designed for 3-way (male, female, or brand related) classification of users; a \tool{} variant supports gender classification.

To ensure generality, we conducted an empirical study to compare \tool{} and its variant with other approaches, using two third-party datasets. 
Our model demonstrated better performance and obtained more balanced results regarding the different roles.

By investigating different classifiers and parameter settings in our hybrid model, we explored how \tool{}'s effectiveness varies with  classifier type, number of tweets, and size of the k-top word set. 
After experimenting with different types and sets of features, we concluded that users' names and profile images are good indicators of the roles of users.
We also observed that female-related users are more likely to post self-related tweets; profile images of brand-related user are brighter. 

In the future, we will investigate the use of more features for classification and apply deep neural networks to enhance feature fusion. 
We plan to transform our model into a hierarchical model, to explore whether prediction results can be improved. 
We also plan to investigate more automatic ways to identify role-related users.